\documentclass[conference,9pt]{IEEEtran}
\IEEEoverridecommandlockouts
\usepackage{subcaption}
\usepackage[ruled,vlined,linesnumbered,noend]{algorithm2e}
\usepackage{algpseudocode}
\usepackage{multirow}
\usepackage{graphicx}
\usepackage{amsmath}
\usepackage{comment}
\usepackage{tablefootnote}
\usepackage[bookmarks=true,breaklinks=true,letterpaper=true,colorlinks,citecolor=blue,linkcolor=blue,urlcolor=blue]{hyperref}

\AtBeginDocument{%
  \providecommand\BibTeX{{%
    \normalfont B\kern-0.5em{\scshape i\kern-0.25em b}\kern-0.8em\TeX}}}

\usepackage{amssymb}
\usepackage{xcolor}
\newboolean{showcomments}
\setboolean{showcomments}{true}
\ifthenelse{\boolean{showcomments}}
{ \newcommand{\mynote}[3]{
		\fbox{\bfseries\sffamily\scriptsize#1}
		{\small$\blacktriangleright$\textsf{\emph{\color{#3}{#2}}}$\blacktriangleleft$}}}
{ \newcommand{\mynote}[3]{}}
\newcommand{\shrink}[1]{}
\definecolor{pink}{rgb}{1,0.2,0.7}
\definecolor{purple}{rgb}{0.7,0,0.9}

\usepackage{tikz}
\newcommand\encircle[1]{%
  \tikz[baseline=(X.base)] 
    \node (X) [draw, font=\small, align=center, shape=circle, inner sep=-0.2ex, scale=1, fill=black, text=white] {\strut #1};%
}
\algnewcommand\algorithmicforeach{\textbf{for each}}
\algdef{S}[FOR]{ForEach}[1]{\algorithmicforeach\ #1\ \algorithmicdo}

\begin{document}

\title{ReCross: Efficient Embedding Reduction Scheme for In-Memory Computing using ReRAM-Based Crossbar}

\author{
\IEEEauthorblockN{
Yu-Hong Lai\IEEEauthorrefmark{1},
Chieh-Lin Tsai\IEEEauthorrefmark{1},
Wen Sheng Lim\IEEEauthorrefmark{1},
Han-Wen Hu\IEEEauthorrefmark{2},
Tei-Wei Kuo\IEEEauthorrefmark{1}\IEEEauthorrefmark{3},
Yuan-Hao Chang\IEEEauthorrefmark{1}
} \\
\IEEEauthorblockA{\IEEEauthorrefmark{1}National Taiwan University, Taipei, Taiwan \\
\{R10922104, d09922013, d08922028, ktw, johnson\}@csie.ntu.edu.tw}
\IEEEauthorblockA{\IEEEauthorrefmark{2}Macronix International Co., Ltd., Hsinchu, Taiwan \\
paddingtonhu@mxic.com.tw}
\IEEEauthorblockA{\IEEEauthorrefmark{3}Delta Electronics, Taipei, Taiwan}
}

\maketitle
\pagestyle{plain}
\begin{abstract}
    Deep learning-based recommendation models (DLRMs) are widely deployed in commercial applications to enhance user experience. However, the large and sparse embedding layers in these models impose substantial memory bandwidth bottlenecks due to high memory access costs and irregular access patterns, leading to increased inference time and energy consumption. While resistive random access memory (ReRAM) based crossbars offer a fast and energy-efficient solution through in-memory embedding reduction operations, naïvely mapping embeddings onto crossbar arrays leads to poor crossbar utilization and thus degrades performance. We present ReCross, an efficient ReRAM-based in-memory computing (IMC) scheme designed to minimize execution time and enhance energy efficiency in DLRM embedding reduction. ReCross co-optimizes embedding access patterns and ReRAM crossbar characteristics by intelligently grouping and mapping co-occurring embeddings, replicating frequently accessed embeddings across crossbars, and dynamically selecting in-memory processing operations using a newly designed dynamic switch ADC circuit that considers runtime energy trade-offs. Experimental results demonstrate that ReCross achieves a 3.97× reduction in execution time and a 6.1× improvement in energy efficiency compared to state-of-the-art IMC approaches.
\end{abstract}

\section{Introduction}

Deep learning-based recommendation models (DLRMs) have become widely adopted across various domains, offering personalized recommendations by analyzing user-item interactions \cite{covington2016deep, DLRM, DBLP:journals/corr/ChengKHSCAACCIA16}. These models enhance user engagement and improve content delivery, significantly benefiting applications such as online browsing and digital advertising. By tailoring recommendations to individual preferences, DLRMs not only enhance user experience but also drive revenue growth in industries reliant on targeted content and advertisements.

A typical DLRM, as illustrated in \figurename~\ref{fig:embedding_lookup}, consists of two multi-layer perceptron (MLP) layers and an embedding layer. The embedding layer maps low-dimensional categorical features to high-dimensional dense vectors, known as embeddings, which serve as inputs to the top MLP layer. This process involves numerous irregular memory lookups~\cite{flashembedding}. These embedding tables are typically stored in DRAM, with the CPU handling embedding lookup operations~\cite{gupta2020deeprecsys,9065589}. However, the large memory requirements and frequent data transfers between memory and CPU introduce significant performance bottlenecks\footnote{Irregular embedding accesses due to categorical input features in embedding lookups pose challenges for data buffering and prefetching in memory.}. For instance, Meta's data center DLRMs require up to 4 terabytes of embedding tables to be loaded into memory~\cite{lui2021understanding}. This scale demands substantial energy consumption and expensive DRAM hardware to support real-time personalized recommendations. Without efficient memory management, off-chip data movement becomes a primary bottleneck, severely impacting inference time and energy efficiency~\cite{9065589, recnmp}. Studies indicate that embedding lookups alone contribute to 50\%–75\% of the runtime overhead in DLRM inference~\cite{merci, gupta2020architectural}.

In-memory computing (IMC) redefines the traditional von Neumann architecture by enabling computations directly within memory, thereby reducing data transfer overhead between memory and processing units~\cite{IMC, IMC2}. Among various IMC technologies, resistive random access memory (ReRAM) crossbar arrays have emerged as a promising solution~\cite{wong2012metal, memristor, VLSI, switching}, where memory cell resistance encodes weights for neural networks or matrix operations~\cite{reno, training}. By applying voltages across the crossbar, matrix multiplications can be executed efficiently, leveraging the inherent parallelism of the ReRAM architecture~\cite{PRIME, isaac}.

The IMC architecture presents a significant opportunity to enhance the performance of memory-bounded applications~\cite{wong2015memory}, particularly for DLRMs~\cite{rerec, imars, nmars}. While most prior studies on IMC-based DLRM acceleration have focused on computing the MLP layers due to their intrinsic multiply-and-accumulate (MAC) operations, embedding reduction can also be transformed into MAC operations. By replacing traditional table lookup operations with in-memory MAC computations, ReRAM-based crossbar arrays can be leveraged to accelerate embedding layers, offering a promising approach to further enhance DLRM performance.

Nonetheless, directly applying in-memory computing to the embedding layer may not yield the expected performance benefits due to its irregular access patterns~\cite{ko2022survey, recssd}. Inefficient embedding placement can cause embeddings from the same query to be dispersed across multiple crossbars, following a power-law distribution, leading to low crossbar utilization and incurring higher latency and energy consumption.

This work is motivated by our key findings that the irregular access behavior of both the embedding access frequency and the co-occurrence of embeddings (i.e., embeddings that strongly correlate with each other) cannot be fully resolved through a straightforward grouping strategy. First, we observe that some crossbars are accessed by multiple queries simultaneously, causing later queries to experience long delays while waiting for prior queries to complete. Second, we observe that most crossbar require only single bitline activation due to the sparsity of embeddings and their irregular access behavior. This results in inefficient use of the multiply-and-accumulate (MAC) operation, leading to unnecessary time and energy consumption for hardware activation. To overcome these challenges, we propose ReCross, an efficient embedding reduction scheme that optimizes in-memory computing for the embedding layer in DLRMs by maximizing crossbar utilization. The key contributions of ReCross include:
\begin{itemize}
    \item \textit{Correlation-aware embedding grouping}, which leverages embedding access patterns and correlation to construct a novel co-occurrence graph that optimally groups embeddings while considering ReRAM crossbar configurations, reducing the total number of crossbar activations for each inference.
    \item \textit{Access-aware crossbar allocation}, which enhances crossbar parallelism and minimizes stall cycles by strategically duplicating embeddings across crossbars using logarithmic scaling.
    \item \textit{Energy-aware dynamic switching}, which improves execution time and energy efficiency by incorporating a newly designed dynamic switch ADC that adaptively selects between read and MAC operations based on runtime energy trade-offs.
\end{itemize}

ReCross is implemented and evaluated with NeuroSIM framework~\cite{neurosim} with Amazon Review dataset~\cite{Amazon}. Specifically, the ReRAM crossbar arrays with 64×64 dimensions and 6-bit resolution self-designed dynamic switch ADCs. Experimental results demonstrate that ReCross achieves a 5.2× (3.97×) speedup in execution time and an 8.4× (6.1×) improvement in energy efficiency compared to the baseline nMARs approach. Additionally, ReCross exhibits at least two orders of magnitude higher energy efficiency than conventional CPU and GPU platforms.

\section{Background and Motivation}

\subsection{Embedding Reduction in DLRM} 
Embedding lookup is a fundamental operation in DLRMs~\cite{covington2016deep, DLRM} that extracts categorical features from embedding tables for each recommendation request, capturing user-item interactions and preferences by aggregating and combining user and item embeddings into a single vector. Specifically, these tables are organized such that each row is a unique embedding vector typically comprising several (e.g., 16, 32, or 64) learned features. As illustrated in \figurename~\ref{fig:embedding_lookup}, DLRM retrieves the corresponding embedding vectors stored in an embedding table and aggregates them using a summation operation\footnote{The sum operation is the most commonly adopted implementations in the previous studies and is supported by most of the popular ML/AI frameworks like Tensorflow, PyTorch, and Caffe2.}. This process, known as \textit{embedding reduction}~\cite{merci}, is a well-known process to constitute multiple representative features that are extremely sparse~\cite{merci} into a single unified and categorical embedding vector. 

\subsection{Computing Embedding Reduction in ReRAM-based Crossbar} \label{sec:back_imc}

\begin{figure}
  \centering
  \begin{subfigure}[b]{0.52\linewidth}
    \includegraphics[width=\linewidth]{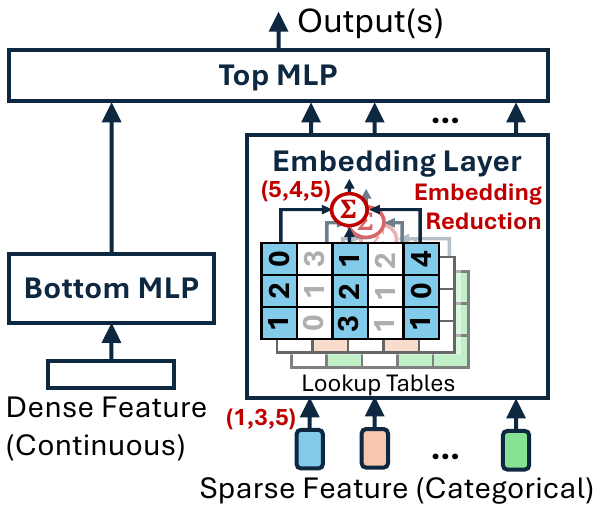}
    \caption{DLRM w/ Embedding Reduction.}
    \label{fig:embedding_lookup}
  \end{subfigure}
   \hfill
  \begin{subfigure}[b]{0.45\linewidth}
    \includegraphics[width=\linewidth]{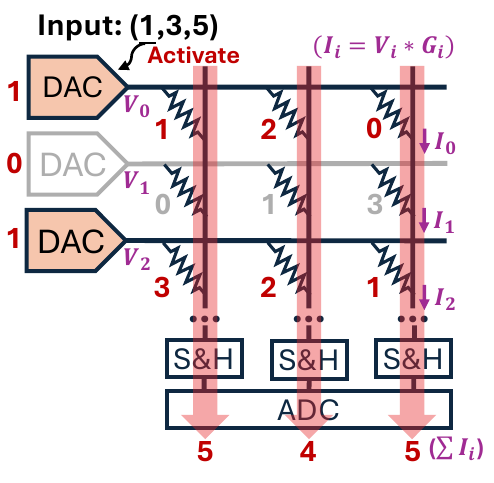}
    \caption{ReRAM-based Crossbar.}
    \label{fig:xbar}
  \end{subfigure}
  \caption{Embedding lookup in DLRM using ReRAM-based in-memory computing architecture.}
  \label{fig:crossbar}
\end{figure}

ReRAM-based crossbar arrays~\cite{wong2012metal} are a well-established in-memory computing (IMC) technology that enables \textit{dot product} operations with low energy consumption while minimizing data movement between memory units and processing elements (e.g., CPU and GPU)\cite{isaac, PRIME}. As illustrated in \figurename\ref{fig:xbar}, ReRAM-based crossbars can perform embedding reduction (depicted in \figurename~\ref{fig:embedding_lookup}) within a single cycle, delivering high computational performance and energy efficiency. In these crossbars, the total current flowing through the bitlines (vertical lines) represents the dot product result, obtained by summing the currents from individual memory cells in a column ($\sum{I_i}$). This structure enables parallel dot product computations across $m$-entry vectors for $n$ distinct neurons. By storing weights as ReRAM conductance values and applying electrical inputs ($V$) on the wordlines (horizontal lines), the embedding reduction can be efficiently transformed into a highly parallel Multiplication and Accumulation (MAC) operation. However, the analog nature of computations in crossbar arrays necessitates the integration of an analog-to-digital converter (ADC) to convert analog currents (MAC results) into digital signals. Notably, the ADC is one of the most power-intensive components in crossbar-based architectures~\cite{isaac}.

\subsection{Characteristic Analysis of DLRM's Embedding Lookup}
\begin{figure}[htb]
  \centering
  \begin{subfigure}[b]{0.46\linewidth}
    \includegraphics[width=\linewidth]{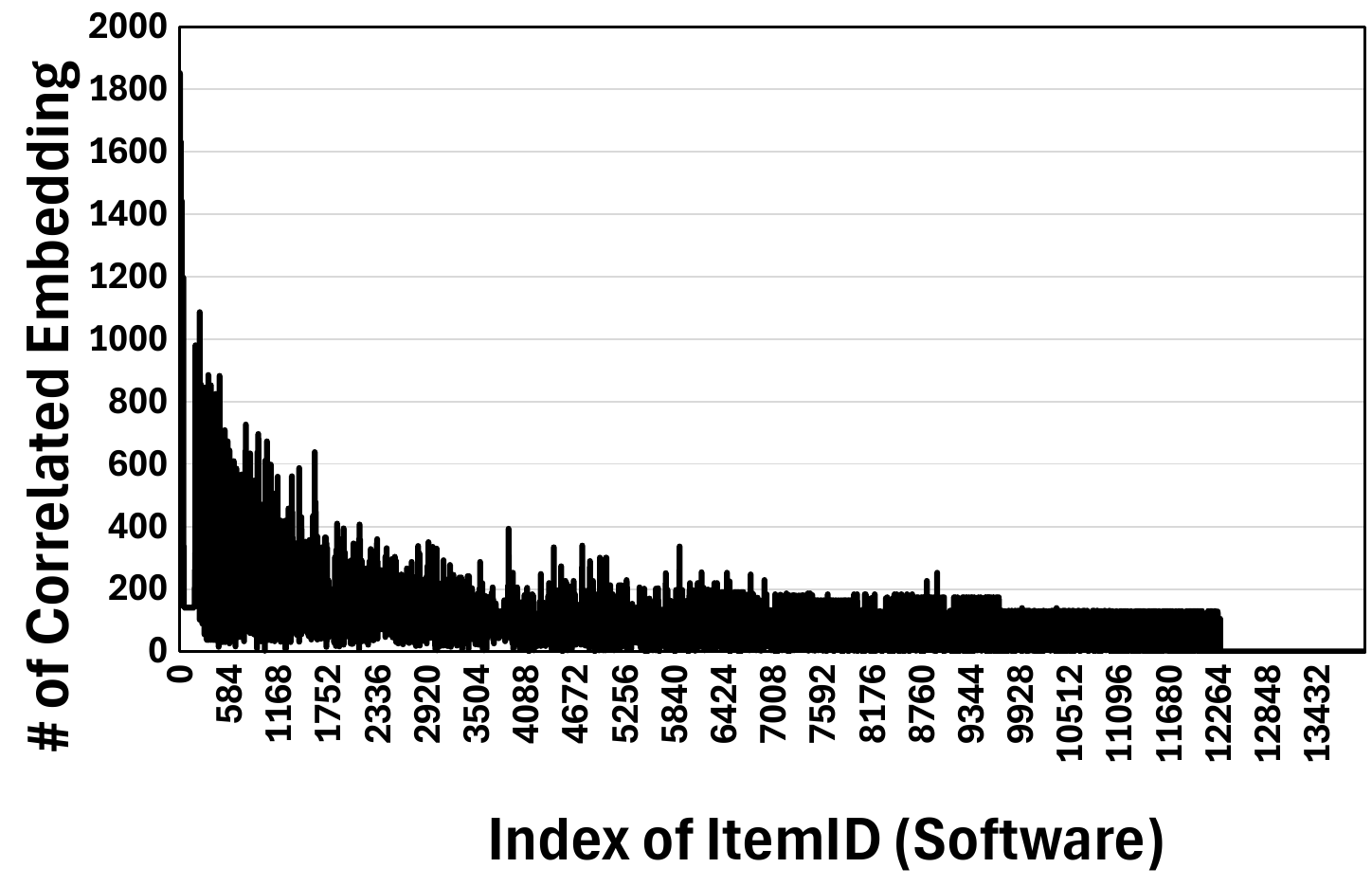}
    \caption{Software.}
    \label{fig:obs1}
  \end{subfigure}
   \hfill
  \begin{subfigure}[b]{0.49\linewidth}
    \includegraphics[width=\linewidth]{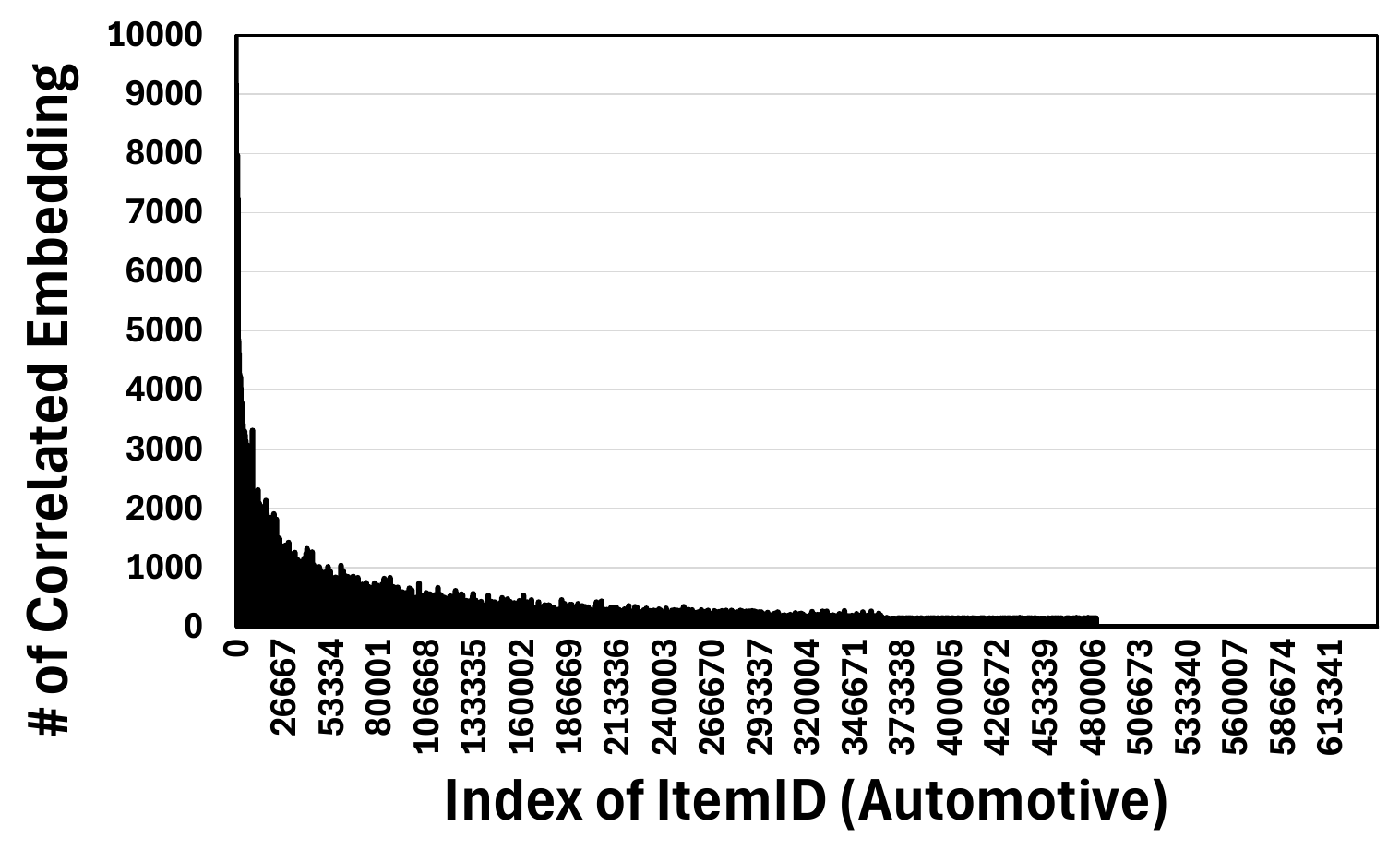}
    \caption{Automotive.}
    \label{fig:obs2}
  \end{subfigure}
  \caption{The number of correlation embeddings.}
  \label{fig:obs}
\end{figure}

Since embedding vectors must be written into ReRAM before processing, the \textit{embeddings-to-crossbar} mapping plays a crucial role in determining the computing bandwidth of the ReRAM crossbar. Efficient mapping ensures fewer crossbar activations (higher utilization), leading to better parallelism and improved inference performance. To analyze embedding lookup behavior, we conducted extensive real-system experiments using the Amazon Review dataset\cite{dataset}. We assign a correlation between two embeddings if they are accessed together, as illustrated in \figurename\ref{fig:obs}. Our findings reveal that both embedding access frequency and co-occurrence follow a power-law distribution, indicating that frequently accessed embeddings are more likely to be strongly correlated with others. In other words, these embeddings correspond to popular or commonly occurring items in the dataset and are frequently involved in diverse contexts and interactions.

\subsection{Motivation and Challenge}
Based on the characteristics of DLRM's embedding lookup, naïve mapping and allocation of crossbar resources can result in low crossbar utilization, particularly when embeddings from the same query are distributed across multiple crossbars (e.g., only 2 out of 3 crossbars are utilized in \figurename~\ref{fig:xbar}). This inefficiency increases the number of crossbar activations, leading to higher completion time and energy consumption during embedding reduction. Additionally, since a single embedding can be accessed by multiple queries concurrently, crossbars storing frequently accessed embeddings may become performance bottlenecks, struggling to process multiple queries simultaneously. Without optimized embedding-to-crossbar placement, crossbar adoption may fail to accelerate embedding reduction and could even degrade the efficiency of in-memory computing.

We present ReCross, an efficient embedding reduction scheme designed to eliminate the data movement bottleneck between the CPU and memory by fully leveraging the computational capabilities of the ReRAM crossbar. The core idea of ReCross is to intelligently \textit{group and map} embeddings onto the crossbar while considering both access frequency and co-occurrence relationships. Through our embedding grouping algorithm, we identify two key challenges. First, the power-law distribution of embedding access patterns \textit{persists} even after grouping, leading to performance degradation due to query delays. To address this challenge, we introduce a crossbar allocation strategy that carefully \textit{duplicates frequently accessed embeddings} across multiple crossbars to minimize contention. Second, we observe that a significant portion of queries require only a \textit{single} embedding access, activating just one row of the crossbar. Since this operation is equivalent to a simple read, performing a full multiply-and-accumulate (MAC) operation results in unnecessary computational overhead, primarily due to the excessive activation of ADCs. To address this challenge, we propose a redesigned ADC circuit that dynamically \textit{switches between read and MAC operations} , reducing unnecessary computations and energy consumption during embedding reduction.

\section{The Proposed ReCross}

\begin{figure}
    \centering
    \includegraphics[width=0.45\textwidth]{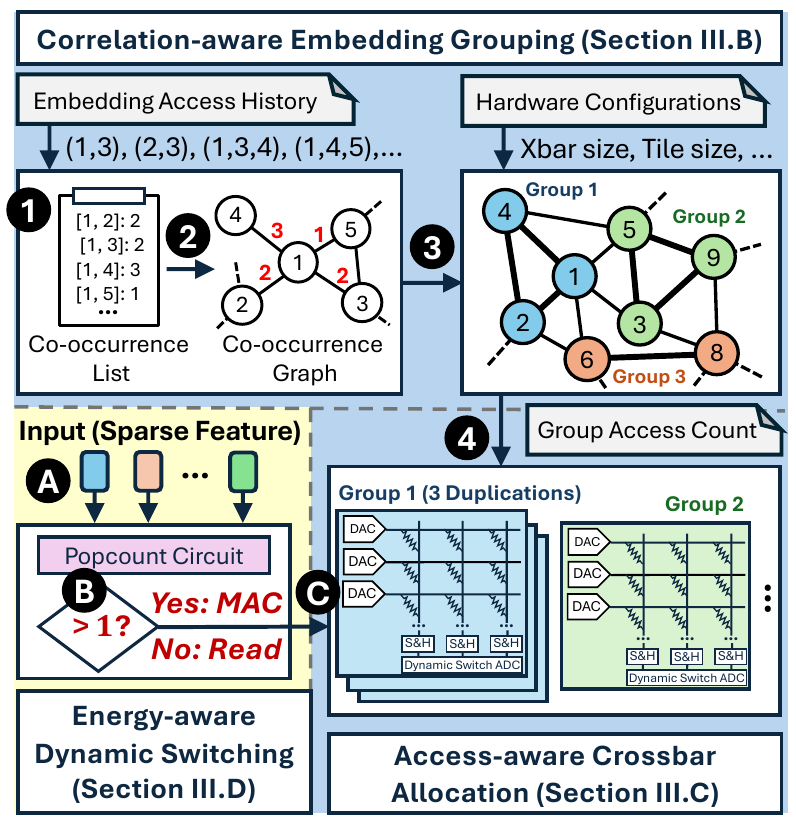}
    \caption{An overview of ReCross system architecture which consists of an online phase (components within blue block) and an online phase (component within yellow block).}
    \label{fig:Overview}
\end{figure}

\subsection{Overview} \label{sec:overview}
ReCross is designed to enable a fast and energy-efficient embedding layer in DLRM systems by performing embedding reduction within \underline{Re}RAM-based \underline{cross}bar in-memory computing architectures. It fully exploits inference parallelism (in batch) while eliminating the data movement bottleneck between the CPU and memory. The primary goal of ReCross is to minimize completion time and maximize energy efficiency in DLRM embedding lookups by reducing the number of crossbar activations while maximizing crossbar utilization for each inference.

\textbf{System architecture.} ReCross consists of two phase: (1) an \textit{offline} which efficiently allocates embedding reduction across crossbar arrays by analyzing workload access patterns and crossbar configurations, and (2) an \textit{online} phase, which dynamically determines crossbar operations at runtime by optimizing the energy trade-off between read and MAC operations. \figurename~\ref{fig:Overview} provides an overview of the ReCross system architecture. In the offline phase, ReCross \encircle{1} constructs a co-occurrence list by analyzing embedding lookup history and tracking the correlation between embeddings based on their access frequency. This list is then \encircle{2} transformed into a co-occurrence graph, where edge weights represent co-occurrence frequency, enabling crossbar activation reduction and improved utilization. Using this graph, ReCross determines the embedding-to-crossbar mapping by \encircle{3} grouping embeddings based on co-occurrence strength (Section~\ref{sec:grouping}) and \encircle{4} further optimizing placement according to access frequency (Section~\ref{sec:duplicate}). Before inference, the embedding table is preloaded into ReRAM based on this optimized mapping. During the online phase, for each \encircle{A} input query, ReCross \encircle{B} dynamically selects the appropriate crossbar operation for embedding reduction. This decision is guided by \encircle{C} an energy-aware trade-off mechanism, leveraging the dynamic switch ADC and popcount circuit to adaptively choose between read and MAC operations (Section~\ref{sec:dynamic_switch}).

\subsection{Correlation-aware Embedding Grouping} \label{sec:grouping}
The correlation-aware embedding grouping strategy constructs a co-occurrence graph using a co-occurrence list, derived from the historical access frequency of embedding lookups. In this graph, nodes represent embeddings, and edges indicate co-occurrence relationships, with edge weights reflecting the frequency of co-access. This structure provides valuable insights into which embeddings frequently appear together, guiding an optimized grouping process. To enhance efficiency, embeddings with stronger co-occurrence (higher edge weights) are prioritized for grouping. By merging frequently co-accessed embeddings into the same group, ReCross reduces the number of crossbar activations and improves overall utilization. Instead of activating multiple crossbars separately for each embedding, ReCross activates a single crossbar containing a cluster of correlated embeddings, minimizing redundant operations. Additionally, edges connected to merged embeddings are preserved to maintain structural integrity.

\begin{algorithm}[ht]
    \scriptsize
    \caption{Correlation-aware Embedding Grouping}
    \label{CAG}
    
    \KwIn{embeddingList, groupSize}
    \KwOut{groupedEmbeddings}

    \textbf{Initialize:} groupedEmbeddings, groupedIndices, candidateList, currentGroup;
    \;
    \ForEach{embedding $\in$ sorted(embeddingList)}{
        \If{embedding $\in$ groupedIndices}{ 
            \textbf{continue}\\
        }
        \eIf{candidateList is empty}{ 
            candidateList $\gets$ neighbors(embedding)\\
        }{
            candidateList $\gets$ Merge(candidateList, neighbors(embedding))\\
        }
        \vspace{3pt}
        maxWeight $\gets$ -1, maxWeightEmbedding $\gets$ None\\
        \vspace{3pt}
        \ForEach{currentEmbedding $\in$ candidateList}{ 
            edgeWeight $\gets$ ComputeWeight(embedding, currentEmbedding)\\
            \If{edgeWeight $>$ maxWeight}{ 
                maxWeight $\gets$ edgeWeight\\
                maxWeightEmbedding $\gets$ currentEmbedding\\
            }
        }
        \vspace{3pt}
        Append(currentGroup, maxWeightEmbedding)\\
        AddToSet(groupedIndices, maxWeightEmbedding)\\
        candidateList $\gets$ Merge(candidateList, neighbors(maxWeightEmbedding))\\        
        \vspace{3pt}
        \If{Size(currentGroup) = groupSize}{ 
            Append(groupedEmbeddings, currentGroup)\\
            currentGroup $\gets$ []\;
        }
    }
    \vspace{3pt}
    \Return groupedEmbeddings\;
    
\end{algorithm}

Algorithm~\ref{CAG} details the correlation-aware grouping process. The algorithm iterates through the sorted embedding list, skipping embeddings that have already been grouped (lines 3–5). It then identifies the embedding in the \texttt{candidateList} with the highest co-occurrence weight to ensure effective grouping (lines 6–13). The selected embedding is added to the \texttt{currentGroup} list, and the \texttt{candidateList} is updated accordingly (lines 14–16). This process continues until all embeddings are grouped or the desired group size is reached (lines 17–19). The final grouped embeddings are recorded in the \texttt{groupedEmbeddings} list. By leveraging this algorithm, ReCross significantly improves embedding layer efficiency, reducing crossbar activations and optimizing resource usage, which is critical for enhancing recommendation system performance.

\subsection{Access-aware Crossbar Allocation} \label{sec:duplicate}
\begin{figure}[h]
  \centering
  \begin{subfigure}[b]{0.48\linewidth}
    \includegraphics[width=\linewidth]{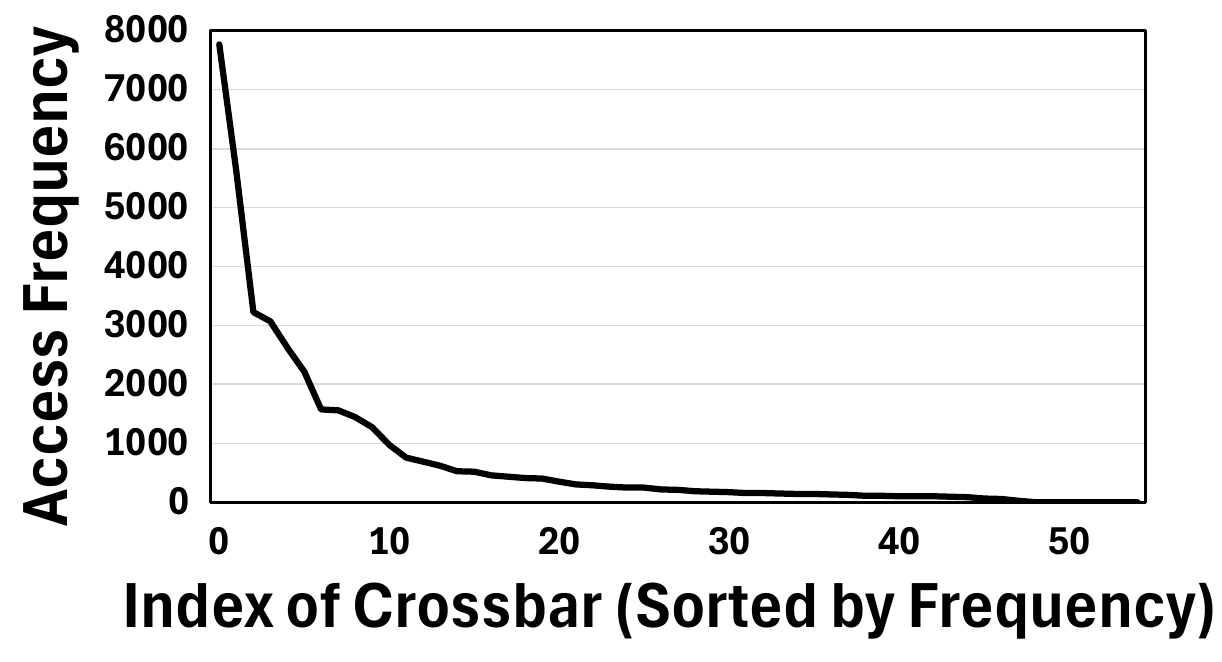}
    \caption{Software.}
    \label{fig:after_group_1}
  \end{subfigure}
   \hfill
  \begin{subfigure}[b]{0.48\linewidth}
    \includegraphics[width=\linewidth]{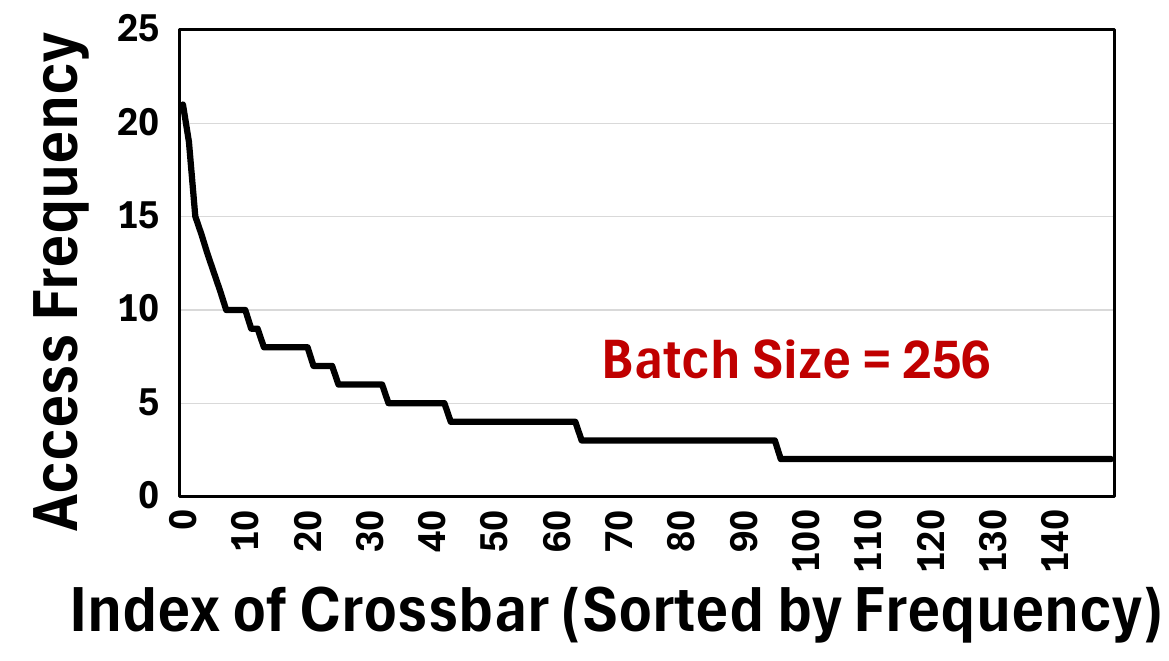}
    \caption{Software with batch of 256.}
    \label{fig:after_group_batch}
  \end{subfigure}
  \caption{Accessed distribution after performing correlation-aware embedding grouping strategy.}
  \label{fig:after_group}
\end{figure}

Our preliminary experiments revealed that even after grouping related embeddings within a crossbar array, the power-law distribution of embedding accesses persists, whether for single or batch-level accesses\footnote{Batch-level inferences are widely used in modern deep learning to maximize parallelism in computing units such as GPUs and crossbar arrays.}, as shown in \figurename~\ref{fig:after_group}. This phenomenon indicates that a small subset of crossbar arrays is accessed disproportionately often, while the majority remain underutilized. Consequently, frequently accessed crossbars dominate the embedded reduction process, limiting parallelism and reducing overall efficiency.

\begin{figure}[h]
    \centering
    \includegraphics[width=0.7\linewidth]{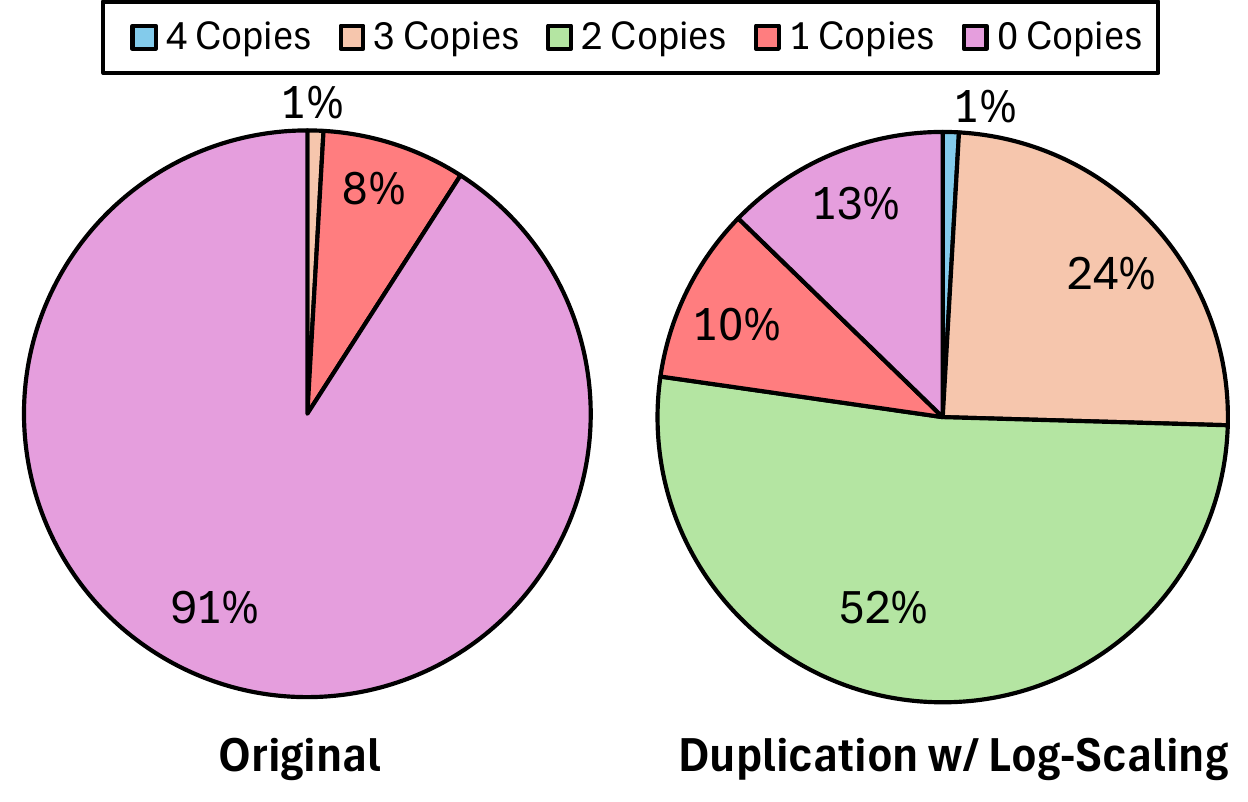}
    \caption{Distribution on the number of copies before and after applying log-scaling in access-aware crossbar allocation.} 
    \label{fig:log_copy}
\end{figure}

To address this issue, we propose to duplicating frequently accessed crossbar arrays to enhance resource utilization and computational parallelism. However, naïve duplication based on original access frequency is suboptimal. As shown in the left pie chart of \figurename~\ref{fig:log_copy}, most crossbars remain unduplicated due to the power-law distribution, where only a small fraction of crossbars experience significantly higher access rates. Additionally, real-world workloads exhibit low query overlap across crossbars. \figurename~\ref{fig:after_group_batch}, which illustrates access distribution for the automotive dataset with a batch size of 256, shows that the maximum access frequency is only 21, far below the batch size. This observation suggests that duplicating embeddings excessively per crossbar is unnecessary, as it does not align with actual access patterns.

To this end, we propose to apply \textit{log scaling} to the crossbar frequencies before determining the number of copies. Log scaling serves two key purposes. First, it mitigates the impact of extreme values inherent in power-law distributions, preventing excessive duplication of embeddings while ensuring that even less frequently accessed vectors receive an appropriate number of copies. Second, it enables a more balanced distribution of duplicated embeddings across crossbars, optimizing computational resource utilization. The number of copies assigned to each embedding is determined by the following equation.
\begin{equation} \label{eqn:duplication}
\small
{Num}_{{copies}} = \left\lfloor \frac{\log({freq})}{\log({freq}_{{total}})} \times \log({batch}_{{size}}) \right\rfloor
\end{equation}
where the ${freq}$ is the frequency of the target embedding, the ${freq}_{{total}}$ is the frequency of all the embeddings, and the ${batch}_{{size}}$ is the available batch size for the duplication. The right pie chart in \figurename~\ref{fig:log_copy} demonstrates the evenness of applying log scaling,  resulting in a better distribution of copies among crossbar arrays.

\subsection{Energy-aware Dynamic Switching} \label{sec:dynamic_switch}
\begin{figure}[h]
  \centering
  \begin{subfigure}{0.48\linewidth}
    \includegraphics[width=\linewidth]{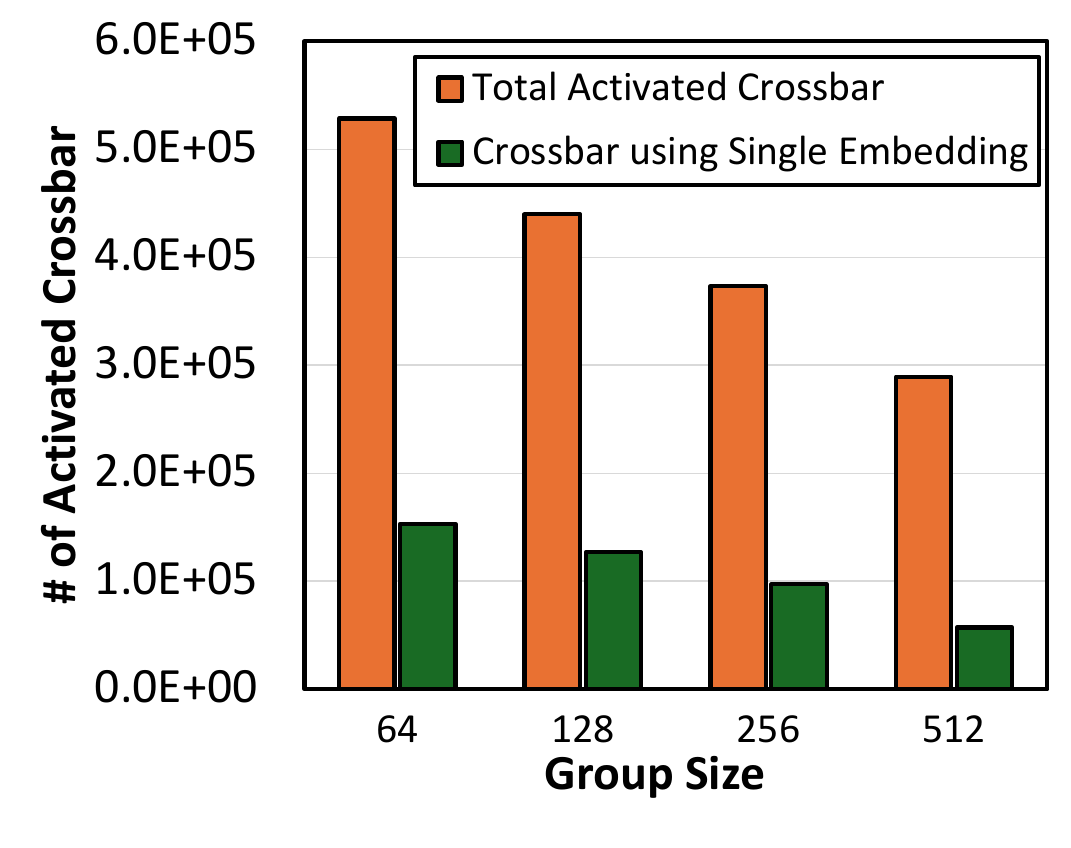}
    \caption{Software.}
    \label{fig:subfig1}
  \end{subfigure}
  \hfill
  \begin{subfigure}{0.49\linewidth}
    \includegraphics[width=\linewidth]{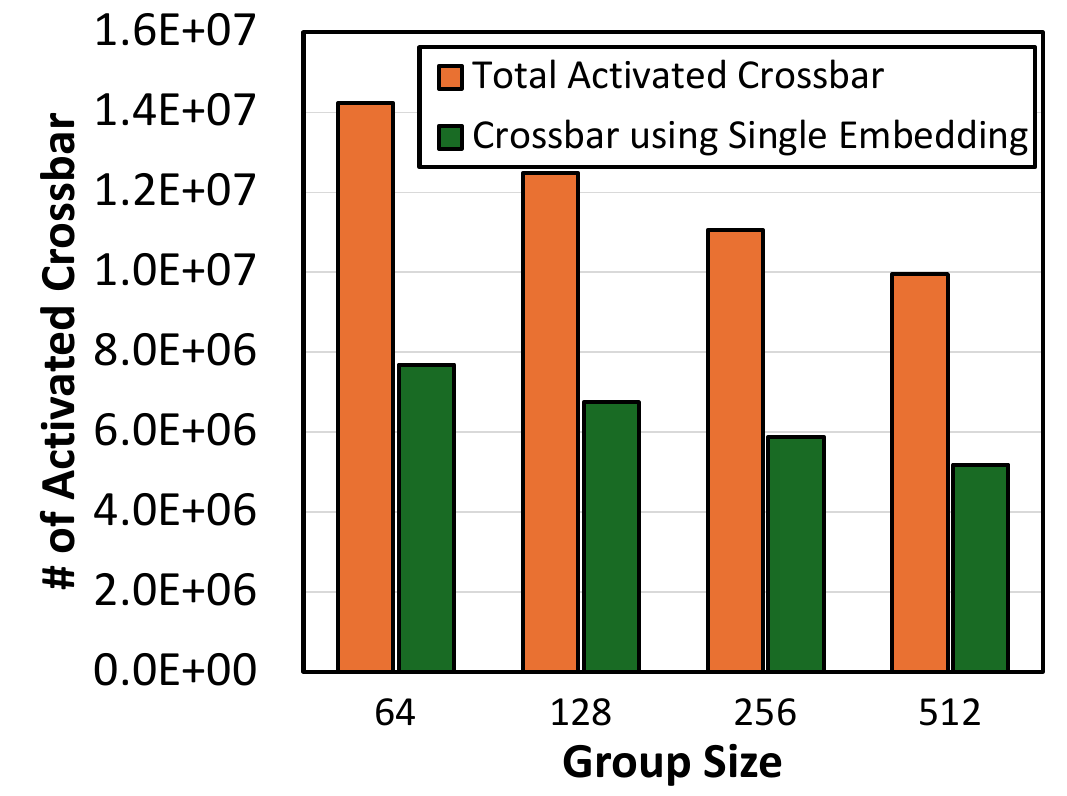}
    \caption{Automotive.}
    \label{fig:subfig2}
  \end{subfigure}
  \caption{The number of activated crossbar with only single embedding access under different group sizes.}
  \label{fig:singleemb}  
\end{figure}

After applying correlation-aware embedding grouping, we observed that a significant portion of crossbars are activated for only a single embedding. As shown in \figurename~\ref{fig:singleemb}, an average of 25.9\% in software workloads and 53.5\% in automotive workloads involve single-embedding accesses. In such cases, the crossbar array activates only one row to retrieve the required data. However, ADCs are designed to performs full-resolution conversion for all analog inputs, regardless of the number of activated crossbars, leading to unnecessary power consumption. For single-embedding retrieval, a read operation is significantly more efficient than a MAC operation, as it eliminates redundant matrix multiplication and accumulation (MAC) computations, particularly in the power-intensive ADC stage~\cite{isaac}.

Based on this insight, we propose a dynamic switching ADC circuit that enables the ReRAM crossbar to dynamically switch between read and MAC operations. This design is built on the Flash Analog-to-Digital Converter (Flash ADC)~\cite{flashADCIntro, ADCsurvey}, which is one of the fastest ADC architectures optimized for low-latency data conversion. Specifically, the Flash ADC employs a linear voltage ladder with comparators to compare analog input signals against predefined reference voltages. The comparator outputs are then processed by a priority encoder, converting them into binary form. Flash ADC is preferred for its fully parallel architecture, enabling instantaneous signal conversion and minimizing latency in large-scale MAC operations for DLRMs. However, its power consumption scales exponentially with resolution, requiring $2^n - 1$ comparators for an N-bit ADC, making it more energy-intensive than other ADC implementations.

\begin{figure}[t]
    \centering
    \includegraphics[width=0.45\textwidth]{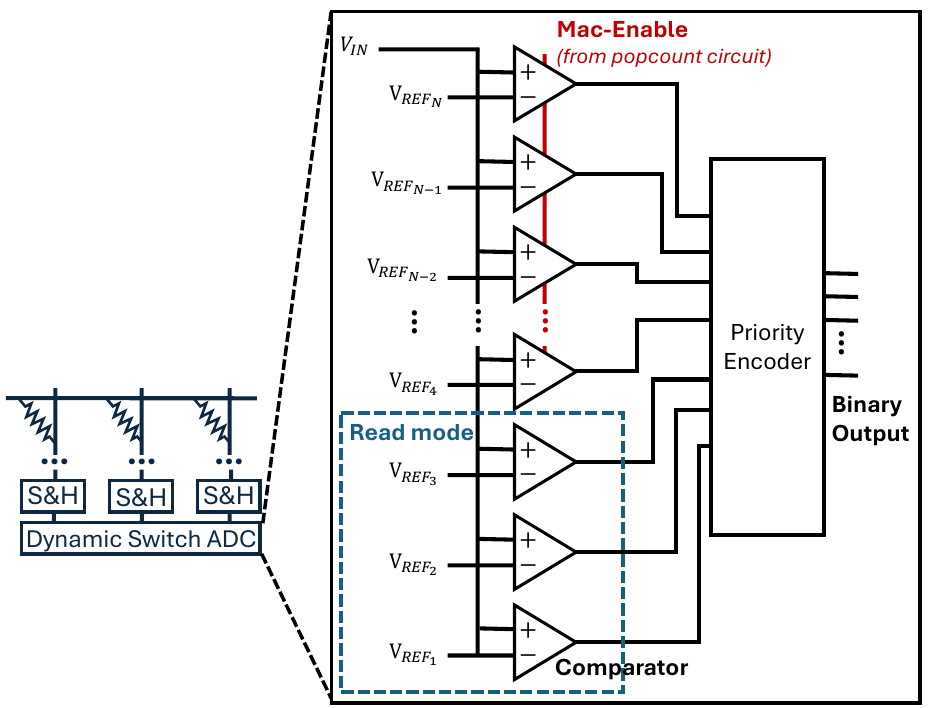}
    \caption{An overview of the proposed dynamic switch ADC.}
    \label{fig:SAR_ADC}
\end{figure}

The dynamic switch ADC aims to leverage between (1) the high occurrence of single-embedding activations in embedding reduction and (2) the excessive power consumption of conventional Flash ADCs. As illustrated in \figurename~\ref{fig:SAR_ADC}, a MAC-enabled signal is added to dynamically control comparator activation, preventing unnecessary circuit operation when processing a single embedding (i.e., only a single row of the crossbar array is activated). When multiple embeddings are accessed, the MAC mode is triggered, fully utilizing the ADC for parallel computation. The switching mechanism relies on a popcount circuit, which efficiently counts the number of '1' bits in the input vector to determine the optimal mode. If multiple '1' bits are detected, the system switches to MAC mode for computation. Conversely, if only a single '1' bit is identified, the system remains in read mode, significantly reducing power consumption while maintaining the high-speed conversion capabilities of the Flash ADC. This adaptive switching approach effectively minimizes power consumption, ensuring that comparator activation dynamically scales with actual computational demand without sacrificing conversion speed. The dynamic switch ADC is lightweight and practical, requiring only minimal hardware modifications, making it an efficient enhancement to conventional ADC designs.

\section{Performance Evaluation} \label{sec:perf_eva}

\subsection{Experiment Setup} \label{sec:setup}
ReCross is evaluated with Neurosim~\cite{neurosim}, a circuit-level simulator, to estimate the latency and energy of the ReRAM-based crossbar array in DLRM embedded reduction based on 22nm technology. The ADC resolution is quantized from 8 bits to 6 bits using NeuroSim's non-linear quantization function, which truncates precision based on the high sparsity of embeddings in recommendation systems. The popcount latency and energy details are referred from~\cite{popcnt}. The evaluation utilizes the Amazon Review dataset~\cite{dataset}, characterized by extensive user-item interactions and diverse patterns common in recommendation scenarios, where five representative workloads are selected according to the diversity of the number of embeddings (i.e., from ~20,000 to ~100,000) to demonstrate the effectiveness of ReCross. All the corresponding hardware configurations and dataset details are outlined in Table~\ref{tab:config}.

\begin{table}[h]
\resizebox{\columnwidth}{!}{
\begin{tabular}{|c|c|l|c|c|c|}
\cline{1-2} \cline{4-6}
\textbf{Component} & \textbf{Specification} & \textbf{} & \textbf{Dataset} & \textbf{\# of Embedding} & \textbf{Avg. Lat} \\ \cline{1-2} \cline{4-6} 
Crossbar & 64 x 64; 2-bit/cell &  & Software & 26,815 & 41.32 \\ \cline{1-2} \cline{4-6} 
Tile & 256 x 256 &  & Office\_Products & 315,644 & 64.088 \\ \cline{1-2} \cline{4-6} 
ADC & 6 bits &  & Electronics & 786,868 & 55.746 \\ \cline{1-2} \cline{4-6} 
Global Bus Width & 512b &  & Automotive & 932,019 & 42.26 \\ \cline{1-2} \cline{4-6} 
Number of Tiles & 4500 &  & Sports & 962,876 & 96.019 \\ \cline{1-2} \cline{4-6} 
\end{tabular}}
\caption{Hardware and dataset configurations.}
\label{tab:config}
\end{table}

\subsection{Experimental Result} \label{sec:result}
The experiments are used to evaluate the performance delivered by ReCross compared with the (1) baseline approach (denoted as naïve) by intuitively mapping the embeddings to crossbar based on the original itemID and (2) nMAR~\cite{nmars} architecture (denoted as nMARS) by considering the conventional embedding reduction in crossbar-based in-memory computing and aggregate them sequentially after parallel in-memory lookup. Two key performance metrics are assessed: (1) average completion time and (2) energy consumption of the embedding lookup. Note that energy efficiency is measured by normalizing the energy consumption of each approach relative to the baseline, with lower energy consumption indicating higher efficiency.

\begin{figure}[tbh]
  \centering
  \begin{subfigure}[b]{0.48\linewidth}
    \includegraphics[width=1\textwidth]{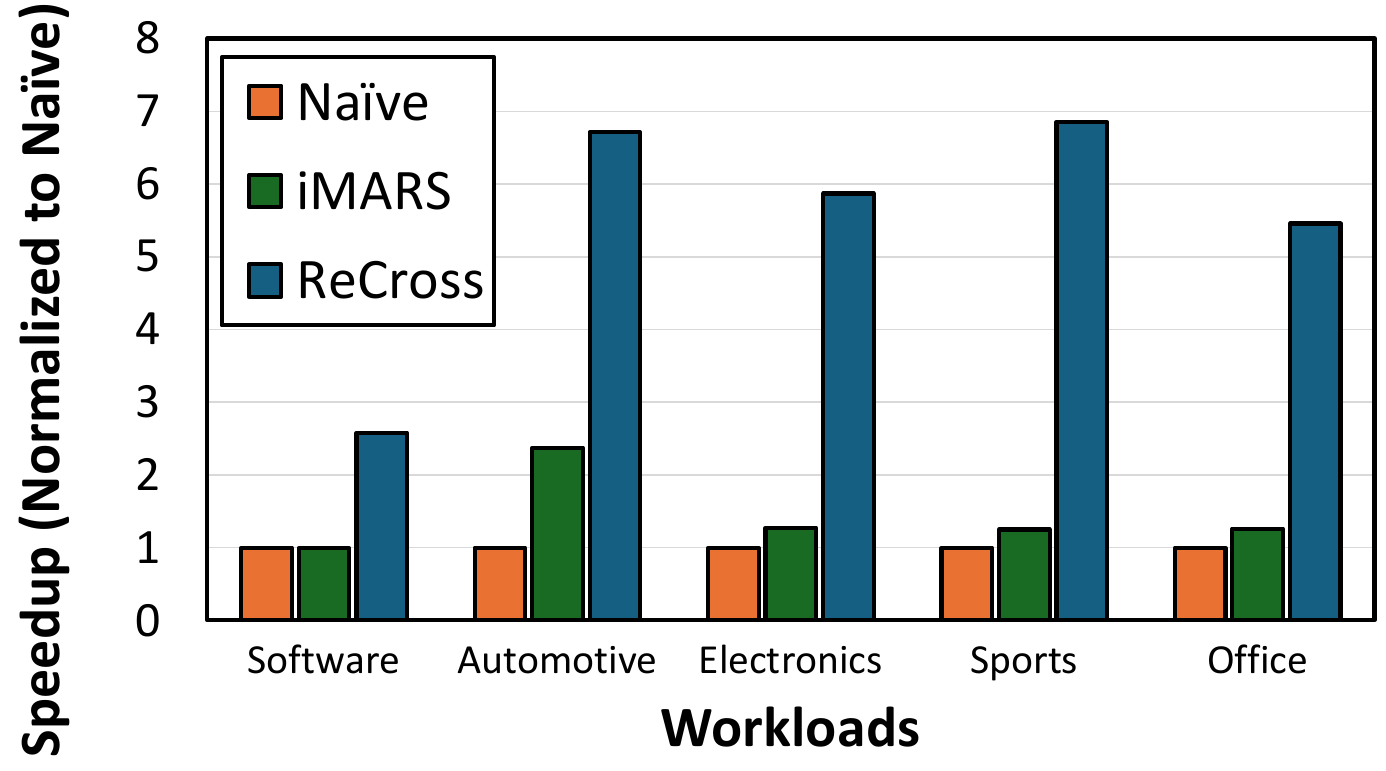}
    \caption{Normalized speedup.}
    \label{fig:subfig1}
  \end{subfigure}
  \hfill
  \begin{subfigure}[b]{0.48\linewidth}
    \includegraphics[width=1\textwidth]{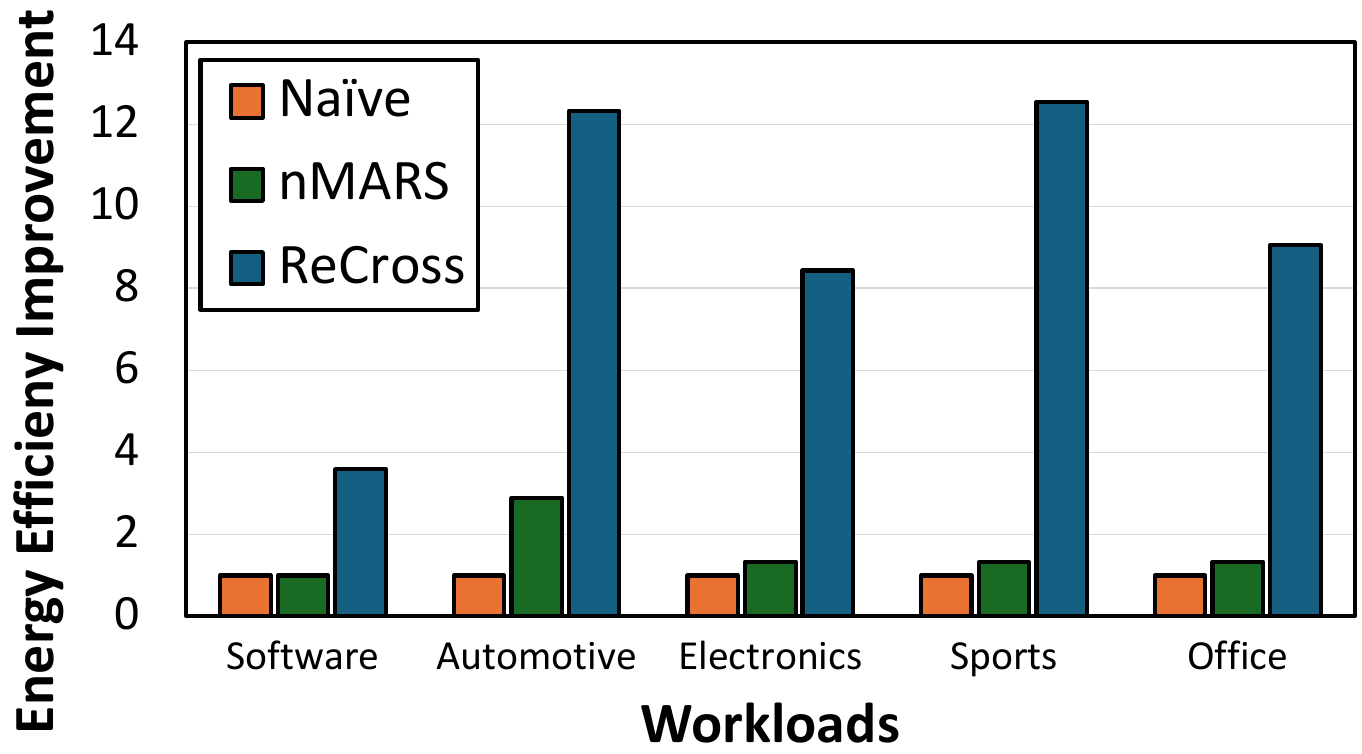}
    \caption{Energy efficiency.}
    \label{fig:subfig2}
  \end{subfigure}
  \caption{Overall performance of ReCross v.s. other approaches.}
  \label{fig:after_exp}
\end{figure}

\textbf{Overall discussion.}
\figurename~\ref{fig:after_exp} compare the normalized execution time and energy efficiency of embedded lookup delivered by ReCross and the other approaches under different workloads. Specifically, ReCross can achieve significantly better execution time by 2.58x (2.60x), 6.71x (2.82x), 5.87x (4.61x), 6.85x (5.48x), and 5.45x (4.33x) and better energy efficiency by 3.60x (1.39x), 12.33x (1.64x), 8.44x (2.43x), 12.55x (3.65x), and 9.06x (2.62x) compared to naive (nMARS) in all workloads, respectively. The major reason is that ReCross is able to minimize the number of crossbar activations (maximize the crossbar resource utilization) by considering the correlation between embeddings and by efficiently duplicating the embedding reduction operations to multiple crossbar arrays. Furthermore, ReCross shows its effectiveness on energy efficiency by leveraging the dynamic switching capability of the ReRAM crossbar based on the observation that a large portion of recommendation inferences require only a single embedding. Specifically, the implemented 6-bit ADCs can reduce energy consumption by 100\% per ADC activation for MAC operations when single embedding is required (i.e., read mode) by utilizing only 3 bits instead of the full 6-bit resolution.

\begin{figure}[tbh]
    \centering
    \includegraphics[width=0.45\textwidth]{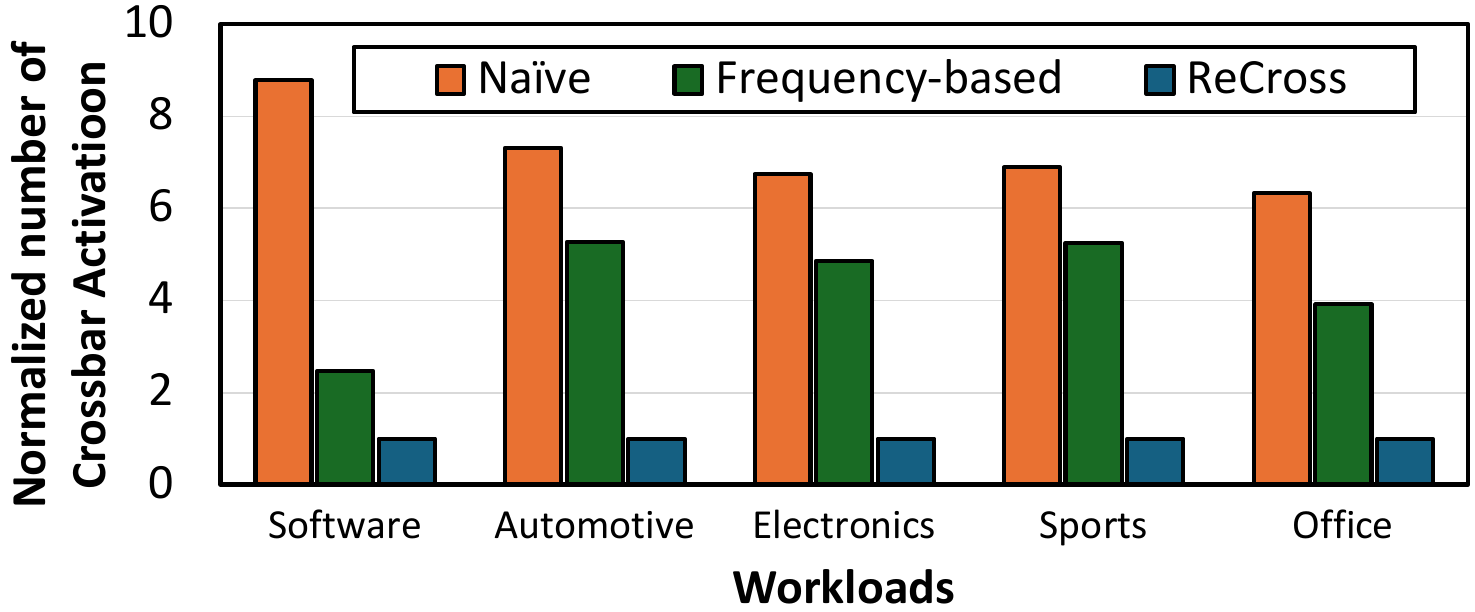}
    \caption{The number of crossbar activations of ReCross v.s. other approaches.}
    \label{fig:crossbar_activations}
\end{figure}

\textbf{Effectiveness analysis of correlation-aware embedding grouping.}
\figurename~\ref{fig:crossbar_activations} compare the number of crossbar activation of ReCross compared to naive and frequency-based approach~\cite{wan2022compute}. As a result, ReCross can significantly reduce the number of crossbar activations up to 8.79x and 5.27x with the exact same hardware configurations under all workloads, compared to naive and frequency-based approaches, respectively. This significant activation reduction laid the foundation for the ReCross effectiveness, further explaining the stunning performance in both execution time and energy efficiency illustrated in~\figurename~\ref{fig:after_exp}. 

\begin{figure}[h]
\centering
\begin{subfigure}[b]{0.48\linewidth}
\includegraphics[width=\linewidth]{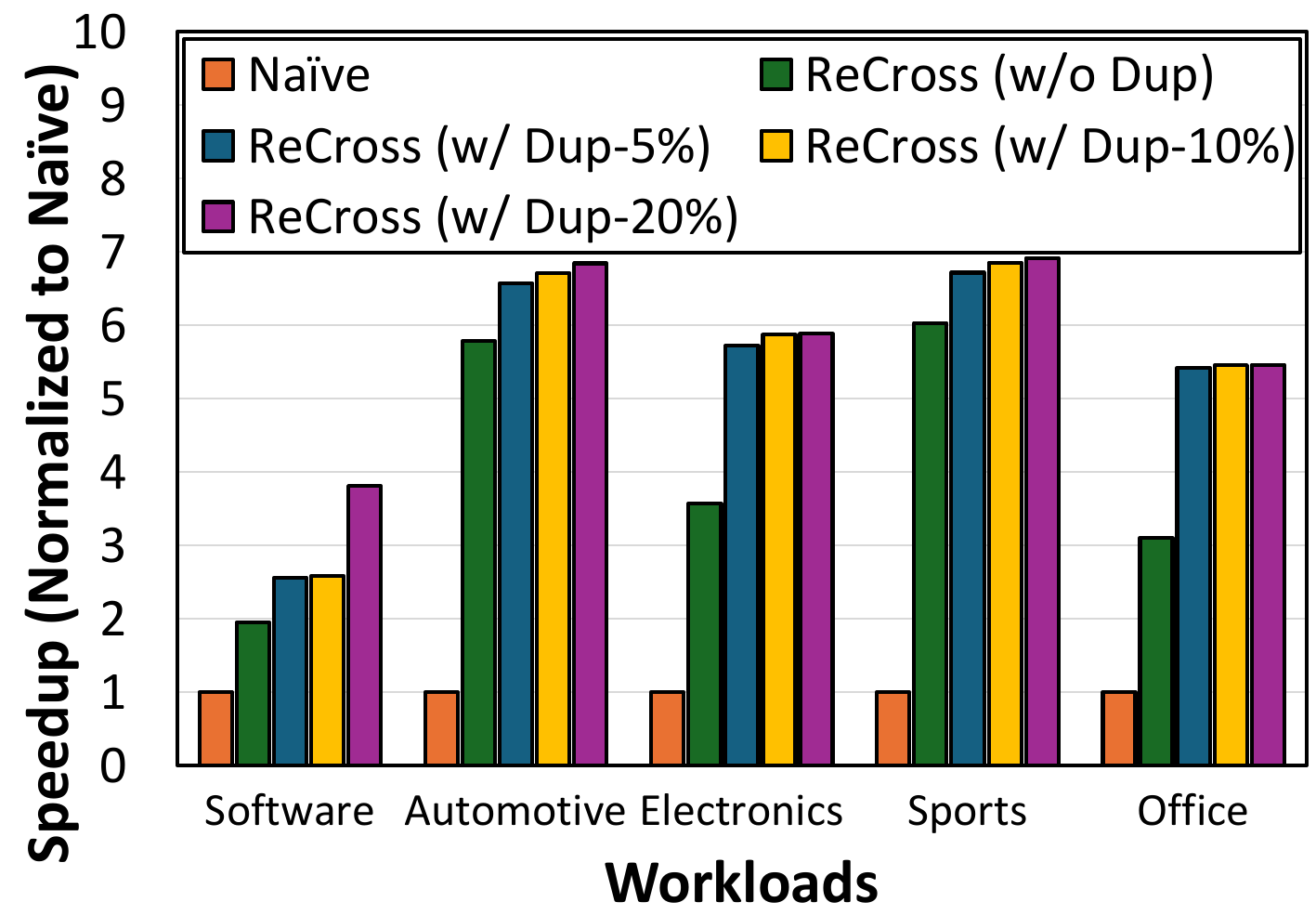}
\caption{Speedup.}
\label{fig:subfig1}
\end{subfigure}
\hfill
\begin{subfigure}[b]{0.48\linewidth}
\includegraphics[width=\linewidth]{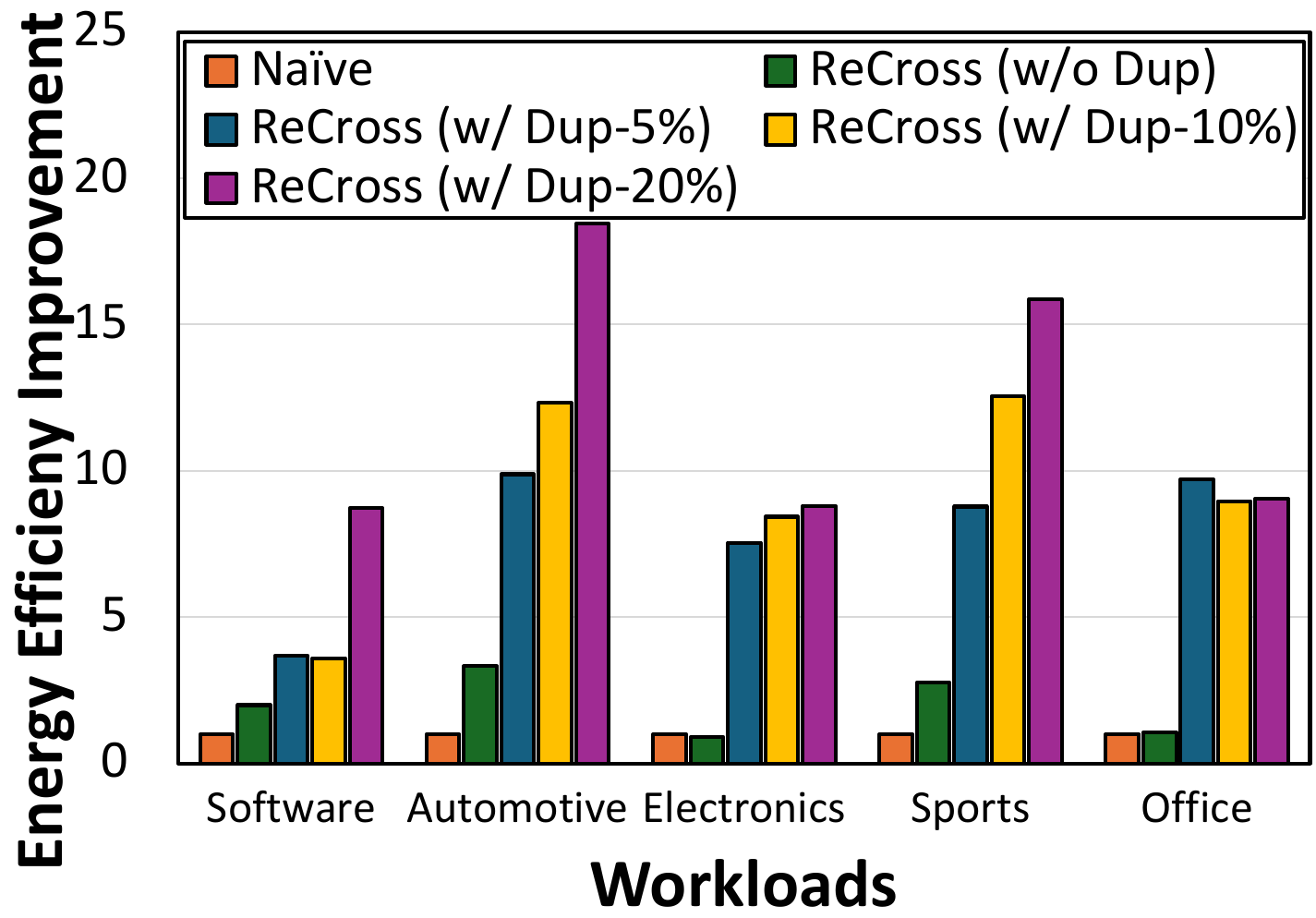}
\caption{Energy Efficiency.}
\label{fig:subfig2}
\end{subfigure}
\caption{Analysis of access-aware crossbar allocation.}
\label{fig:aad_exp}
\end{figure}

\textbf{Effectiveness analysis of access-aware crossbar allocation and corresponding area overhead.}
\figurename~\ref{fig:aad_exp} analyzes the effectiveness of access-aware crossbar allocation that duplicates the frequently accessed embedding to multiple crossbars with different duplication ratios (i.e., 0\%, 5\%, 10\%, and 20\%). These different duplication ratios impose the percentage of extra area overhead of ReCross compared to the baseline approaches. In particular, the proposed crossbar allocation strategy shows its effectiveness in both execution time and energy efficiency compared to the naive approach and simplified version of ReCross (i.e., w/o duplication). Note that the performance improvement of both execution time and energy efficiency starts to converge as the number of duplications increases because the embedding accesses have their limitations, as mentioned in Section~\ref{sec:duplicate}. However, workloads with more dense access behavior (i.e., requiring more duplication for the frequently accessed groups) still show their effectiveness even in Dup-20\% configurations, posing a design trade-off between performance and ReRAM area overhead under various scenarios. We carefully argue that this different performance profiling under different workloads and crossbar configurations indicates a research opportunity for the community to further explore the performance of embedded reduction for a certain classification of workloads in DLRMs.

\begin{figure}[h]
    \centering
    \includegraphics[width=0.42\textwidth]{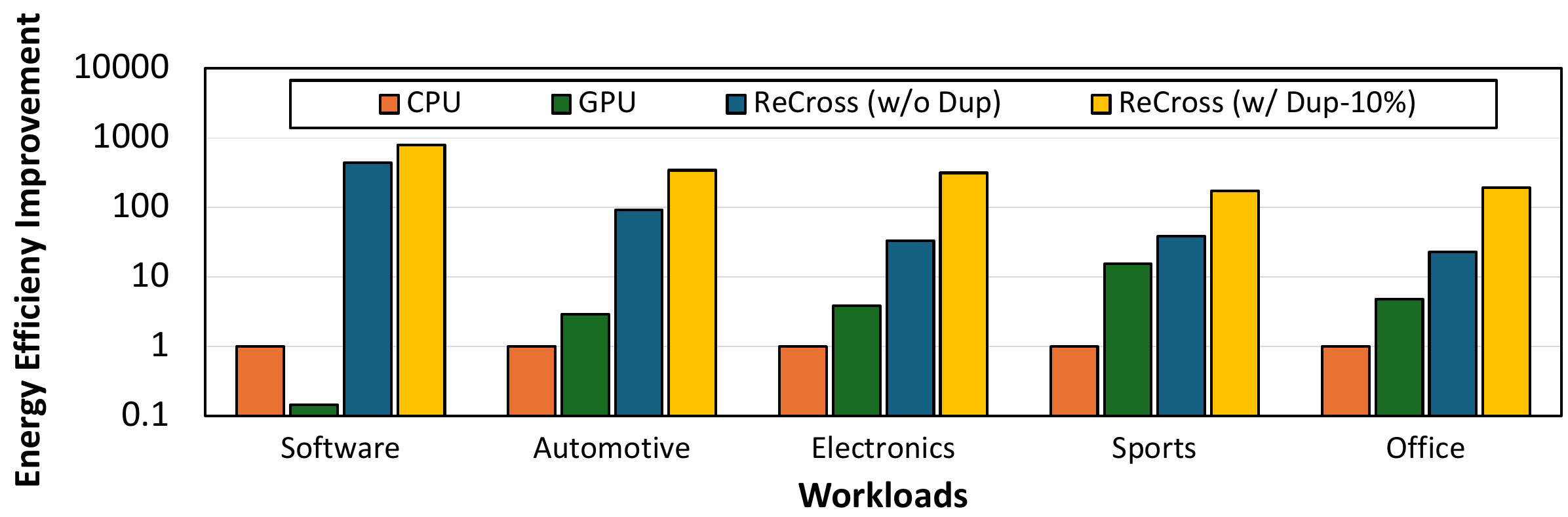}
    \caption{Energy efficiency of ReCross against CPU and GPU.}
    \label{fig:CPUGPU}
\end{figure}

\textbf{Energy efficiency comparison against CPU and GPU.}
\figurename~\ref{fig:CPUGPU} compare the energy efficiency of ReCross against CPU-only architecture and CPU-GPU architecture. The experiments were conducted on a GPU NVIDIA GeForce RTX 3090 and a CPU i7-10700F. The evaluation of energy consumption of CPU and GPU is conducted using the energy profiler provided by MERCI~\cite{merci} and NVIDIA Management Library~\cite{NVML}. Thanks to the ReRAM crossbar architecture, ReCross can impressively outperform on average of 363x and 1144.15x compared to CPU-only architecture and CPU-GPU architecture, respectively. The exaggerated improvement on energy efficiency shows the promising possibilities of in-memory computing architecture against the conventional von Neumann architecture.

\section{Related Works} \label{sec:related_work}
Recent studies have identified embedding reduction as a critical performance bottleneck in recommendation systems, spurring various acceleration approaches through both software optimizations and hardware architectures. On the software optimization front, MERCI~\cite{merci} proposes sub-query memoization to reduce memory accesses by identifying and storing partial sums of frequently co-occurring embeddings, achieving significant speedup without hardware modifications. GRACE~\cite{grace} extends this concept by introducing a graph-based clustering algorithm to identify optimal embedding combinations and optimize their placement in CPU and GPU memory. However, without in-memory MAC support, these approaches must pre-store all combinations and their outcomes, leading to significant memory overhead. Several solutions have explored near- and in-memory computing schemes to address the memory bandwidth bottleneck. 

RecNMP~\cite{recnmp} and TensorDIMM~\cite{tensordimm} integrate computation units near memory banks to reduce data movement, though they still require embedding vectors to be fetched before computation. REREC~\cite{rerec} eliminates data movement by implementing specialized ReRAM-based inner-product engines with small crossbars, managing fine-grained computations through access-aware mapping. nMARS~\cite{imars,nmars} presents a hierarchical in-memory computing fabric that combines crossbars with configurable memory arrays to accelerate recommendation filtering and ranking. However, these approaches primarily focus on conventional embedding lookups without fully leveraging crossbar-based computation capabilities. To this end, our proposed ReCross advances the state-of-the-art by applying in-memory MAC operations for embedding reduction while considering both inter-embedding correlations and in-memory hardware characteristics, further utilizing the parallelism capabilities of ReRAM crossbar for efficient embedding reduction.

\textbf{Differences against ReCross}. Differences Against ReCross. Previous works have primarily focused on either software-level optimizations for embedding reduction or hardware architectures for ReRAM crossbars. In contrast, ReCross is the first to explore the acceleration of DLRM embedding reduction using ReRAM crossbars, integrating both data-aware optimizations and in-memory computing strategies to maximize efficiency.

\section{Conclusion}
The efficient embedding reduction scheme ReCross is proposed to address the challenges of low performance and inefficient resource utilization in ReRAM-based crossbar systems for DLRMs. Specifically, ReCross efficiently groups and maps the co-occurrence embeddings to the ReRAM crossbar, smartly duplicating the frequently access crossbar and dynamically deciding the runtime in-memory processing operations based on the newly designed dynamnic switch ADC. Our results demonstrate ReCross outperforms the prior in-memory computing architecture nMARs by 3.97x and 2.35x on avearage, in execution time and energy efficiency, respectively. ReCross also demonstrates at least two orders of magnitude in energy efficiency with superior speedup compared to CPU and GPU platforms, showcasing the efficacy of in-memory computing in significantly enhancing the performance of such memory-bounded applications.

\bibliographystyle{IEEEtran}
\bibliography{reference}

\end{document}